\documentclass[twocolumn,prl,aps,showpacs]{revtex4}
\usepackage{graphicx}

\begin{document}

\title{Comment on ``Intermittent Synchronization in a Pair of Coupled
Chaotic Pendula"}
\author{P.~Muruganandam}
\author{S.~Parthasarathy}
\author{M.~Lakshmanan}
\affiliation{Centre for Nonlinear Dynamics, Department of Physics, 
Bharathidasan University, Tiruchirapalli 620 024, India}

\maketitle

In a recent Letter, Baker, Blackburn and Smith~\cite{baker} reported
that permanent synchronization of a pair of unidirectionally coupled
identical chaotic pendula does not occur except as a numerical artifact
arising from finite computational precisions. Baker {\em et al.} showed
that the synchronization of a pair of unidirectionally coupled pendula
is always {\em intermittent}, for any value of coupling coefficient
$c$, by using their numerical and analytical tests. This was also found
to be true in the case of  coupled Duffing oscillators~\cite{baker}.
However, on careful numerical analysis using Conditional Lyapunov
Exponents (CLEs), we find that there exists some specific range of $c$
values for which persistent synchronization can occur for these
systems.

The main point of this comment is to clarify the conditions under which
coupled chaotic systems can exhibit permanent synchronization.
According to Pecora and Carroll\cite{pecora}, the master and slave
systems will perfectly synchronize only if the sub-Lyapunov exponents
or CLEs are all negative. The two analytic criteria mentioned in
Ref.\ \cite{baker} for asymptotic stability are, ({\em i}) the largest
eigenvalue of the Jacobian matrix corresponding to the flow evaluated
on the synchronization manifold must always be negative and ({\em ii})
existence of a suitable Lyapunov function.  However,
it is not possible, in general, to show analytically the above two
criteria are uniformly obeyed for all $c$ values for most of the
chaotic systems, as shown in the case of coupled pendula (cf.
Eqs.\ (9) and (11) of Ref.\ \cite{baker}).  In such cases, it is always
possible to compute the Lyapunov exponents of the coupled system for a
range of $c$ values and show that the CLEs take large negative values
for permanent synchronization.  

\begin{figure}[!ht]
\begin{center}
\includegraphics[width=\linewidth]{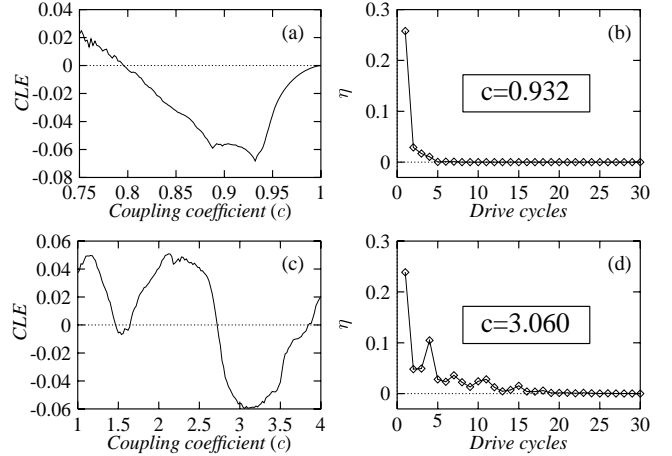}
\end{center}
\caption{
(a) Variation of the Conditional Lyapunov Exponent (CLE) as a function
of $c$, (b) synchronization error $\eta$ (given by Eq.\ (5) in Ref.~[1])
versus time in drive cycles for coupled pendula; (c) CLE versus $c$ and
(d) $\eta$ versus drive cycles for coupled Duffing oscillators.
}
\label{fig1}
\end{figure}

We have computed the Lyapunov exponents for the coupled pendula and the
Duffing oscillators using the standard Wolf {\it et al.} algorithm.
For numerical simulations we have used the same parameter values as in
Ref.\ \cite{baker} and used $5000$ drive cycles for calculations after
leaving $5000$ drive cycles as transient.  Figure \ref{fig1}(a) shows
the variation of CLE as a function of $c$ for the coupled pendula and
it takes negative values for $0.796 \le c < 1.0$ only. The CLE value
for $c=0.79$ is $+0.00320$ ($\approx 10^{-3}$), for which
synchronization can not occur and hence the observed intermittent
synchronization (cf. Fig.\ 1 in Ref.\ \cite{baker}) is a computer
artifact.  However, we find that for $c=0.932$, the CLE value becomes
the lowest ($-0.06825$) and is a large negative value for which
intermittent synchronization is absent as shown in Fig.\ \ref{fig1}(b).
This has been verified upto $2\times 10^6$ drive cycles with the
addition of tiny noise levels showing that permanent synchronization
does occur. The same phenomenon persists over a range of $c$ values
close to $0.932$.

Similar arguments hold good for the coupled Duffing oscillator also and
the results are shown in Figs.\ \ref{fig1}(c) and (d). The CLE takes
small negative values close to zero ($\approx 10^{-3}$) for
$c\in(1.48,1.64)$ where synchronization is intermittent and numerical
artifact arises in this range also. The CLE value for $c=1.5$ is
$-0.00562$ and hence the observed hard locking becomes intermittent
with the addition of noise in the $8$-th digit (cf. Fig.~4 in
Ref.~\cite{baker}). In addition, there exists another range of $c$
values, $2.74\le c\le 3.84$ where the CLEs are negative (cf.
Fig.\ \ref{fig1}(c)). We find that for $c=3.06$, the CLE takes the
lowest value of $-0.05967$ (large negative value) for which persistent
synchronization occurs and the intermittent synchronization is absent
(cf. Fig.\ \ref{fig1}(d)). Again the same phenomenon persists over
a range of $c$ values close to $3.06$.

In summary, the main aim of our comment is to emphasize that the
conditional Lyapunov exponents play an important role in distinguishing
between intermittent and persistent synchronization, when it is not
possible to show analytically the criteria for asymptotic stability are
uniformly obeyed. We find that intermittent synchronization can occur
when CLEs are very small positive or negative values close to zero
while persistent synchronization occurs when CLEs become sufficiently
large negative values.

This work was supported by DST, Govt. of India (M.L. and S.P.),
DAE-NBHM (M.L.) and CSIR (P.M.).


\begin{thebibliography}{99}
\bibitem{baker}
G.~L.~Baker, J.~A.~Blackburn and H.~J.~T.~Smith, \prl
{\bf 81}, 554 (1998)
\bibitem{pecora}
L.~M.~Pecora and T.~L.~Carroll, \prl
{\bf 64}, 821 (1990)
\end{thebibliography}
\end{document}